\documentstyle[titlepage,fleqn,12pt]{article}
\begin{document}

\begin{center}
{\Large \bf{Stochastics theory of log-periodic patterns}}

\vspace{1cm}
{\bf Enrique Canessa}\footnote{E-mail: canessae@ictp.trieste.it} \\
{\em The Abdus Salam International Centre for Theoretical Physics,
Trieste, Italy}
\end{center}

\vspace{2cm}
{\baselineskip=20pt
\begin{center}
{\bf Abstract}
\end{center}

We introduce an analytical model based on birth-death clustering
processes to help understanding the empirical log-periodic
corrections to power-law scaling and the finite-time singularity as
reported in several domains including rupture, earthquakes, world
population and financial systems.  In our stochastics theory
log-periodicities are a consequence of transient clusters induced
by an entropy-like term that may reflect the amount of cooperative
information carried by the state of a large system of different
species.  The clustering completion rates for the system are assumed 
to be given by a simple linear death process.  The singularity at 
$t_{o}$ is derived in terms of birth-death clustering coefficients.

\vspace{1.5cm}
}

\baselineskip=22pt
\parskip=4pt

\pagebreak

\section{Overview}

Increasing evidence of accelerated patterns having an overall power
law behaviour with superimposed log-periodic oscillations has been
found in a variety of applied domains.  These observations have been
reported in a series of experiments on rupture in heterogenous media
\cite{Ani95,Joh98} and from the historical data analysis of earthquakes
\cite{Sor95,Sal96,Joh96,Joh00}, world population \cite{Joh00a} and
financial stock markets \cite{Joh00b,Sor96,Fei96,Glu98,Van98,Dro99},
(see examples in Fig.1).  It has been also argued that log-periodic
corrections to scaling should be present in a wider class of
out-of-equilibrium dynamical systems (see, {\em e.g.}, \cite{Shl91,Sau96}).
The logarithmic modulations are periodic in $t-t_{o}$ and not on $t$ and
are precursors to an spontaneous finite-time singularity $t_{o}$ at which
they accumulate.

The interest in log-periodic corrections to power-law scaling is twofold.
On one hand they enhance the fit quality to observed data with better
precision than simple power laws by adjusting the (frequency, local minima
and maxima of) oscillations.  On the other hand their real-time monitoring
could, in principle, allow for an enhancement of predictions in different
contexts \cite{Sau96,Sor98}.

At the theoretical level, log-periodic oscillatory structures has been
associated to the existence of complex fractal dimensions \cite{Sor96}
and critical exponents \cite{Sau96,Sor98,Sor96a,Sor97}.  However, as
pointed out in \cite{Lal99}, predictions of stock market crashes using
complex critical exponents should be taken with some concern not only
because of the many fitting parameters required but also because the
time period used to perform the fit is rather long \cite{Can00}.  This
does not mean that the apparent acceleration and the log-periodic
modulation do not actually exist -the whole subject deserves further
investigations.

Most recently, log-periodic patterns associated to financial crashes
have also been shown to stem from models for stock markets inspired
by percolation phenomena \cite{Sta98,Sta00,Sta00a}.  Furthermore,
logarithmic oscillations have been found in an off-lattice bead-spring
model of a polymer chain in a quenched porous medium under the
influence of an external field \cite{Yam99} and in a uniform spin
model on a fractal \cite{Les00}.  However, besides these
simulation studies, there is no a convincing microscopic theoretical
model which substantiates the idea that the singularity at finite
$t_{o}$ ({\em e.g.}, a financial crash) is a critical point.  Also,
there is as yet no a fundamental theory that substantiates the claim
for the precursory, universal log-periodic oscillations on large time
scales.

In this work we made an attempt to provide a new scenario within
which to elaborate an analytical microscopic theory and contribute
towards an understanding of the underlying physics of log-periodic
patterns.  We introduce an stochastics model based on birth-death
clustering processes to sustain the claim of log-periodic corrections
to scaling and of a finite-time singularity.  In our theory transient
clusters are formed following an entropy-like formula that may
reflect the amount of cooperative information (or disorder) carried
by the state of a large system of different species.  The clustering
completion rates for the system are assumed to be exponentially
distributed according to a simple linear death process.  The 
singularity at $t_{o}$ is derived in terms of birth-death clustering 
coefficients.

\section{Stochastics theory}

We write the governing equation for power law behaviour decorated
by large scale log-periodic oscillations as a superposition of two terms
\begin{equation}\label{eq:the}
G(t) = G_{0}(t) + G_{\infty}(t) \;\;\; ,
\end{equation}
where $G_{0}$ takes the form of a pure power law and $G_{\infty}$
represents the (universal periodic) corrections.

Similarly to \cite{Ani95,Sor95}, the first term is taken to be
\begin{equation}\label{eq:go}
\frac{dG_{0}(t)}{dt} = \kappa(t_{o}-t)^{\alpha -1} \;\;\; ,
\end{equation}
with $t_{o}$ a finite time at which a singularity appears and the
exponent $\alpha$ satisfies $\alpha \ne 1$.  Integration of this
equation yields
\begin{equation}\label{eq:go1}
G_{0}(t)=G_{0}(t_{o}) - \frac{\kappa}{\alpha}(t_{o}-t)^{\alpha} ;\;\; .
\end{equation}

In the following we seek for an approximate form to the correction
term $G_{\infty}$.

\subsection{Galerkin finite elements method for $G_{N}(s,t)$}

The starting point of our theory is to assume that the two-dimensional
({\em e.g.}, energy and time dependent) $G_{N}$ function of a discretized
system of $N$ nodes is the solution of some non-linear differential equation
({\em e.g.}, a diffusion equation with particular boundary conditions)
which we do not know but we shall figure out the answer.

Using the standard Galerkin finite elements method described in the Appendix
(see also, {\it e.g.}, \cite{Fle84}), a general trial solution to this unknown
differential equation can be approximated as 
\begin{equation}\label{eq:gt}
G_{N}(s,t) \equiv \sum_{j=0}^{N} g_{j}(s) \tilde{P_{j}}(t)  \;\;\; ,
\end{equation}
where $g_{j}$ are basic (interpolation) functions, and $\tilde{P}_{j}$ are
the so-called test functions.  The terms $g_{j}(s)$ are often referred to
as trial functions and Eq.(\ref{eq:gt}) as the trial solution at nodal points.

Without loss of generality, for all $t$ and different $j$-states we can
rewrite $\tilde{P}$ in a more suitable form as the sum of even and odd parts
\begin{equation}\label{eq:pj}
  \tilde{P_{j}}(t) = p_{2j}(t) + p_{2j+1}(t) \;\;\; .
    \;\;\; \;\;\; \;\;\; (j=0,1,2,\cdots,N)
\end{equation}
This means that
$\tilde{P_{0}}=p_{0} + p_{1}$, $\tilde{P_{1}}=p_{2} + p_{3}$,
$\tilde{P_{2}}=p_{4} + p_{5}$, $\cdots$

Hence, in the limit $N\rightarrow \infty$, we then get
\begin{equation}\label{eq:norm}
\sum_{j=0}^{\infty} p_{j}(t) = \sum_{j=0}^{\infty} \tilde{P_{j}}(t)
  \equiv 1 \;\;\; ,
\end{equation}
a result that will prove useful later when we associate
the test functions $\tilde{P}_{j}$ with the equilibrium state
probabilities to be characterized by birth-death processes.

\subsubsection{Birth-death model for $p_{j}(t)$ - Effect of disorder}

The test functions $\tilde{P_{j}}(t)$ are usually determined by
solving a system of differential equations (in time) generated by
some governing equation, and if $N$ is made arbitrarily large the error
introduced becomes small (see Appendix).  In order to gain insight into 
the dynamics leading to log-periodic structures, we take next a different 
approach and relate the test functions to a large number of processes forming
clusters or aggregates that change as a function of time ({\em e.g.}, cell
populations, customers queueing, interactive multi-agent ensembles,
investors groups) acting collectively to pass on information
or to introduce system disorder.

We assume that stochastics "births" and "deaths" clustering processes
occur according to a simple one-dimensional {\em birth-and-death}
model (see, {\em e.g.}, \cite{Goo88}).  The state probabilities $p_{j}(t)$
in this case are obtained recursively from
\begin{equation}\label{eq:rec}
\hat{\lambda}_{j}(t)p_{j}(t) = \hat{\mu}_{j+1}(t)p_{j+1}(t) \;\;\; ;
 \;\;\; \;\;\; \;\;\; (j=0,1,2,\cdots)
\end{equation}
By a choice of the birth coefficients $\hat{\lambda}_{j}>0$
and of the death coefficients $\hat{\mu}_{j}>0$, various stochastics
models can be constructed ({\em e.g.}, queueing models in which
costumers corresponds to the "population", arrivals are "births" and
departures are "deaths").  In other words, the quantity $\hat{\lambda}$
is interpreted as the birth rate and $\hat{\mu}$ the death rate when
the population is at the state $j$.

Eq.(\ref{eq:rec}) together with the normalization condition of
Eq.(\ref{eq:norm}) can easily be solved to yield the following
statistical-equilibrium state distribution (as seen from an
arbitrary outside observer)
\begin{equation}\label{eq:sta}
p_{j}(t) = \frac{\hat{\lambda}_{0}(t)\hat{\lambda}_{1}(t)\cdots
  \hat{\lambda}_{j-1}(t)}
{\hat{\mu}_{1}(t)\hat{\mu}_{2}(t)\cdots\hat{\mu}_{j}(t)} p_{0}(t) \;\;\; .
\end{equation}
This means that for each time $t>0$ the state probabilities can, in
principle, be determined subject to specification of the initial
conditions $p_{0}(t)$ ({\em i.e.}, the so-called absorbing state)
and the product of birth-death ratios at all previous states.

To obtain the above transient solution for $p_{j}(t)$ ({\em i.e.},
for finite $t$) in closed form, it is necessary to postulate basic
expressions for $\hat{\lambda}_{j}$ and $\hat{\mu}_{j}$.  Here we
assume that -for the case of a finite probability distribution,
{\em i.e.}, $p_{j}(t)>0$ (with $j=1,2,\cdots,N$)- clusters form via
an entropy-like formula
\begin{equation}\label{eq:ent}
\hat{\lambda}_{j}(t) =  - \sum_{k=1}^{M}\lambda_{k}(t)
\ln \lambda_{k}(t) > 0 \;\;\; ;  \;\;\; \;\;\; \;\;\; (j=0,1,2,\cdots,N-1)
\end{equation}
with $\lambda >0$ for all $k$.
According to information theory \cite{Ing97}, the shape of our birth
coefficients may reflect the measure of cooperative information carried
by the outcomes $\lambda_{1},\cdots,\lambda_{M}$ (or the amount of disorder
in the discrete observable $\hat{\lambda}_{j}$) in a system of $M$
different species or types ({\em e.g.}, human gender, financial traders).

On the other hand, in analogy to Erlang loss systems \cite{Goo88},
we assume the clustering completion rate for the system in state $j$
to be exponentially distributed (with rate $\mu >0$), hence we deal
with simple linear death processes
\begin{equation}
\hat{\mu}_{j}(t) = j \mu(t) \;\;\; ;
          \;\;\; \;\;\; \;\;\; (j=1,2,\cdots,N)   \;\;\; .
\end{equation}

Then, the equilibrium state probabilities given by Eq.(\ref{eq:sta})
become
\begin{equation}\label{eq:bd}
p_{j}(t) = \frac{(-1)^{j}}{j!} \biggl[ \sum_{k=1}^{M}
 a_{k}\ln \lambda_{k}(t) \biggl]^{j} p_{0}(t)
 \;\;\; , \;\;\; \;\;\; \;\;\; (j=0,1,2,\cdots,N)
\end{equation}
where the {\em per-capita} ratio $a_{k}\equiv \lambda_{k}(t)/\mu(t) >0$
is, for simplicity, assumed to be time independent.  This means that
$\lambda_{k}$ and $\mu$ should both scale as power laws of the form
$\sim \Delta t^{\pm n}$.

By using the normalization condition of Eq.(\ref{eq:norm}) plus 
Eq.(\ref{eq:bd}) and the Taylor series
expansions for the {\em exponential} function, we also obtain
\begin{equation}\label{eq:pt}
p_{0}(t)  \equiv
 \biggl[ \sum_{j=0}^{\infty}\frac{(-1)^{j}}{j!} \Gamma_{M}^{j}(t)
  \biggl]^{-1} = e^{\Gamma_{M}(t)}  \;\;\; ,
\end{equation}
where
\begin{equation}\label{eq:gam}
\Gamma_{M}(t) \equiv \sum_{k=1}^{M} a_{k}\ln \lambda_{k}(t) =
  \ln \prod_{k=1}^{M}\lambda_{k}^{a_{k}}(t) = \ln p_{0}(t)  \;\;\; .
\end{equation}

Now that we have a precise formulation for $\tilde{P_{j}}(t)$,
we can set trial functions for the basic interpolation functions
$g_{j}$ to solve for $G_{N}(s,t)$ given in Eq.(\ref{eq:gt}).

\subsubsection{Trial $g_{j}(s)$ functions}

The efficiency of the Galerkin formulation is very dependent on
making the correct choice of the approximating test and trial functions
(see Appendix).  Of the many nodal unknowns that could be candidates, 
here we consider the polynomial expansion
\begin{equation}\label{eq:gd}
g_{j}(s) \equiv \frac{\gamma}{(s-1)^{j}} = \gamma  \biggl\{  (-1)^{j} +
  \left(  \begin{array}{c}
           j \\ 1
  \end{array} \right)  (-1)^{j-1}s +
  \left(  \begin{array}{c}
           j \\ 2
  \end{array} \right)  (-1)^{j-2}s^{2} +
     \cdots \biggl\}  \;\;\; ,
\end{equation}
with $s$ a dimensionless variable.

Substitution of Eq.(\ref{eq:bd}) into (\ref{eq:pj}) and using
Eq.(\ref{eq:gam}) plus these trial functions, it gives the approximate
trial solution
\begin{equation}\label{eq:gt2}
G_{N}(s,t) =  \gamma p_{0}(t) \sum_{j=0}^{N} \frac{1}{(s-1)^{j}}
  \biggl\{ \frac{(-1)^{2j}}{(2j)!} \Gamma_{M}^{2j}(t) +
\frac{(-1)^{2j+1}}{(2j+1)!} \Gamma_{M}^{2j+1}(t) \biggl\} \;\;\; ,
\end{equation}
as is easily verified.

We shall see next that our $g_{j}$ functions make a judicious choice.

\subsection{Onset of log-periodicity}

Let us consider large systems in statistical equilibrium and adopt the
following notation for the required correction term of Eq.(\ref{eq:the})
near $t_{o}$:
\begin{equation}
G_{\infty}(t) \equiv \lim_{N \rightarrow \infty} G_{N}(s\approx 0,t)
\;\;\; .
\end{equation}
For the sake of simplicity we have set $s\approx 0$ in order to gain
insight into the genesis of log-periodicities.

As an example, our approximate trial solution of Eq.(\ref{eq:gt2}) thus
becomes
\begin{equation}\label{eq:gt3}
G_{\infty}(t) =  \gamma p_{0}(t) \biggl\{
    \sum_{j=0}^{\infty} 
\frac{(-1)^{j}}{(2j)!} \Gamma_{M}^{2j}(t) - \sum_{j=0}^{\infty}
\frac{(-1)^{j}}{(2j+1)!} \Gamma_{M}^{2j+1}(t) \biggl\} \;\;\; .
\end{equation}

Using Taylor series expansions for the {\em cosine} and {\em sine} functions
plus the trigonometric identity $\cos (a+b)=\cos (a)\cos (b)-\sin (a)\sin (b)$,
such that $a=\pi/4$ and $b/2\pi \equiv \Gamma_{M}(t)$ in radians, the above
equation then results in
\begin{equation}\label{eq:gt4}
G_{\infty}(t) = \sqrt{2} \gamma p_{0}(t) \cos \biggl(2\pi \ln p_{0}(t) +
   \frac{\pi}{4}\biggl)  \;\;\; ,
\end{equation}
with $p_{0}$ satisfying Eq.(\ref{eq:gam}).

This equation characterizes the complexity of the underlying dynamics of
the scaling systems under consideration.  It can be seen that in our
stochastics theory the log-periodic modulation is a consequence
of the entropy-like assumption used for the transmission of information
within the birth-death clustering processes.

Using our expression for the initial boundary condition $p_{0}$ leading to
Eq.(\ref{eq:gt4}), we analyse next the presence of a singularity at a finite 
time where the oscillations accumulate.

\subsection{Finite-time singularity}

As discussed in the derivation of Eq.(\ref{eq:bd}) $\mu$ should scale
with $t$, so we set
\begin{equation}\label{eq:mu}
\mu(t) \sim (t_{o} - t)^{\pm n}  \;\;\; ,
\end{equation}
where $n \ne 0$ is a given exponent and $t_{o}$ charaterizes the finite-time
singularty.  Hence, according to our definitions
\begin{equation}\label{eq:lam}
\lambda_{k}(t) \sim a_{k} (t_{o} - t)^{\pm n} \;\;\;  ,
\end{equation}
which implies that within our stochastics birth-death theory we can consider 
finite values of time $t < t_{o}$ (or $n$ even only if $t > t_{o}$) since 
the coefficients $\lambda_{k}$ and $\mu_{k}$ (and, therefore, $a_{k}$) are
all positive. 

By substitution of this scaling into Eq.(\ref{eq:gam}), we finally get
\begin{equation}\label{eq:pt1}
p_{0}(t)  =  \prod_{k=1}^{M}\lambda_{k}^{a_{k}}(t) =
   \Theta_{M} \prod_{k=1}^{M}(t_{o}-t)^{\pm n a_{k}} \;\;\; ,
\end{equation}
where
\begin{equation}
\Theta_{M} \equiv \prod_{k=1}^{M}a_{k}^{a_{k}}  \;\;\; .
\end{equation}

Thus from this relation and Eq.(\ref{eq:gt4}), we are able to
derive log-periodic corrections in the form of $G_{\infty}$ for $t<t_{o}$.

\section{Discussion}

Having introduced our theory based on stochastics clustering processes
to describe log-periodic corrections to scaling and a finite-time
singularity at $t_{o}$, we use Eqs.(\ref{eq:the}),
(\ref{eq:go1}), (\ref{eq:gt4}) and (\ref{eq:pt1}) to obtain
\begin{equation}\label{eq:gt5}
G(t) = A + B (t_{o}-t)^{\alpha} + C (t_{o}-t)^{\beta}
    \cos \biggl( 2 \pi \beta \ln (t_{o}-t) + \psi \biggl)  \;\;\; ,
\end{equation}
where $A \equiv G_{0}(t_{o})$, $B \equiv - \kappa/\alpha$, and $\alpha$ 
are parameters relating the pure power law term of the governing
Eq.(\ref{eq:the}) and
\begin{equation}\label{eq:gt6}
C  \equiv  \sqrt{2}\gamma\Theta_{M}  \;\;\; , \;\;\;
\beta  \equiv  \pm n \sum_{k=1}^{M} a_{k}  \;\;\; , \;\;\;
\psi  \equiv   2 \pi \ln \Theta_{M} + \frac{\pi}{4}  \;\;\; ,
\end{equation}
are the parameters of the log-periodic corrections in terms of our
stochastics birth-death model parameters.  The fit of this equation to
historical random data displaying accelerated precursory patterns and
an spontaneous singularity, which indicates the sharp transition to a
new regime, is presented in Fig.1.

As an illustrative example, in Fig.1 we plot the
daily Log(S\&P500) stock index closing values during the years 1982-1988
\cite{Can00}, the estimated 1000-2000 world population by the
U.N. Population Division \cite{Popul} and the sum of seismic activities
measured near the Virgin Islands between Apr. 1979 and Feb.
1980 \cite{Var96}.  Our best fits with Eq.(\ref{eq:gt5}) to these
data sets have been done as follows:

\begin{center}
\begin{tabular}{|l|c|c|c|c|c|c|c|c|} \hline \hline
 & A & B & C & $\alpha$  & $\beta$ & $t_{o}$ &  $\psi$ &  {\small {\it rms}} \\ \hline
S\&P500 index & 1.51 & 4.34 & -0.01 & -0.1 & 1.42 & 1988.62 & 1.85 & 0.053 \\ \hline
World population & 0.25 & 1489.74 & -25.24 & -1.38 & -1.04 & 2054.61 & -6.34 & 0.047 \\ \hline
Seismic activity & -1.38 & 1.18 & -0.04 & -0.74 & -1.25 & 1980.29 & 0.78 & 0.414 \\ \hline \hline
\end{tabular}
\end{center}

Examination of the plotted curves shows that Eq.(\ref{eq:gt5}) can
model log-periodic corrections to the leading scaling behaviour and
a singularity at $t_{o}$ in different applied domains similarly to
the methods inspired by renormalization group theory entailing
complex critical exponents.  If we set $\alpha = \beta = 1/f$ we can
approach the results obtained in \cite{Joh00a,Joh00b,Sor96} where
market crashes, population explosion, and culminating large earthquakes
are viewed as critical points in a system with an underlying discrete
scale invariance.  If we set $\alpha \ne \beta$, our results are
then comparable to those obtained using the more general anzat of two
different exponents as in \cite{Fei96}.  The main difference of our
stochastics theory with respect to the critical exponent approach is
the presence of the exponent $\beta$ also appearing in the argument of 
the cosine function.  Since $n$ and $a_{k}$ are positive then from 
Eq.(\ref{eq:gt6}) we have that $\beta \ne 0$.  Furthermore, in our 
stochastics theory the finite time $t_{o}$ is also determined 
in function of $\beta$ as discussed below.

So far, the fitting in Fig.1 allows us to argue that the apparent logarithmic
periodicities in scaling systems may also be understood within the
context of an stochastics analytical model based on birth-death clustering
processes which is the distinctive feature of our stochastics theory.  We
can interpret the "births" and the "deaths" clusterings in different ways.
For financial systems the "births" and "deaths" may represent, {\em e.g.},
the buyers and the sellers, respectively.  Whereas newborns and deceases
would be in correspondence with the population growth.  Some absorbed-
and released- energy may relate the "births" and "deaths" in the case
of seismic events for finite times $t<t_{o}$.

The recurrence Eqs.(\ref{eq:rec}) are {\em conservation-of-flow}
relations.  That is, the long-run rate at which the system moves
up from state $j$ to $j+1$ equals the rate at which the system
moves down from state $j+1$ to $j$ ({\em i.e.}, rate up = rate down).
Thus, birth-death processes describe the stochastic evolution in
time of a random variable whose values varies ({\em i.e.}, increases
or decreases) by one in a single event (or state) starting from
the absorbing state $p_{o}$.

The spontaneous singularity is here related to the birth-death
coefficients which in turn determinate $p_{o}$, {\em i.e.} the
initial boundary condition at the state $j=0$ via Eq.(\ref{eq:pt}).
It is also important to note that the state distribution coefficient
defined by
\begin{equation}\label{eq:coeff}
- \biggl( \frac{p_{j+1}-p_{j}}{p_{j}} \biggl) =
       1 +  \frac{\Gamma_{M}}{j+1}   \;\;\; ,
\end{equation}
depends on the logarithm of the absorbing state via Eq.(\ref{eq:gam}).

In the absence of a well-defined non-linear dynamical equation governing
log-periodic corrections to power-law scaling, we have adopted
the standard Galerkin finite elements method as the starting point
to search for a general trial solution to this "unknown" differential
equation.  The motivation for our basic interpolation functions $g_{j}(s)$
follows computational finite-element methods which are characterized by
the use of polynomials for the known test functions (obtained from
Eqs.(\ref{eq:pj}) to (\ref{eq:gam})) as well as for the unkown trial
functions of Eq.(\ref{eq:gd}) in subdomains called finite elements 
\cite{Fle84}.  

As discussed in the Appendix, our trial $g_{j}(s)$'s would allow to solve
the matrix equation for the test functions, which have been related to a
large number of (birth-death) processes forming clusters, provided the 
governing equation of the problem would be known.  These forms, using 
$s\approx 0$, were taken for convenience to gain insight into the onset 
of log-periodicities.  If we consider instead small $s << 1$, we would 
obtain rather similar conclusions after some algebra.

We thus argue that log-periodicities are a consequence of transient
clusters induced by the entropy-like term given in Eq.(\ref{eq:ent})
which may reflect the amount of cooperative information carried by
the state of a large system ({\em i.e.}, $N\rightarrow \infty$) of
different species $M$.  Using the definition of the amount of (discrete)
finite information or entropy, it can be shown that the information is
additive under concatenation of independent probabilities as the logarithm
function is.  It has been proved that it is possible to define information
without necessarily using the concept of probability (see, {\em e.g.},
\cite{Ing97}).  We have adopted the latter definition in this work via
Eq.(\ref{eq:ent}) for the birth coefficients.  The clustering completion 
rates for the system are given by a simple linear death process.

The state probablities $p_{j}$ are normalized via Eq.(\ref{eq:norm}).  We may 
also consider that the total sum of outcomes $\lambda_{k}$ is constant for
all time as in the Shannon theory of information \cite{Ing97}.
Therefore, for the whole range $M$ of different species we set 
\begin{equation}
\sum_{k=1}^{M}\lambda_{k}(t)\equiv \xi_{t} > 0 \;\;\; . 
\end{equation}
We can then estimate the finite time $t_{o}$ at which a singularity appears from 
Eqs.(\ref{eq:lam}) and (\ref{eq:gt6}) by considering the initial time $t=0$ to 
thus obtain the relation
\begin{equation}\label{eq:to}
t_{o}^{\pm n} = \frac{\pm n \xi_{o}}{\beta} \;\;\; .
\end{equation}
From the examples in Fig.1 we found for the Log(S\&P500) stock index: 
$n=0.105 $, $\xi_{o}=30$. World population: $n=-0.209$, $\xi_{o}=1$.
Accumulated seismic activity: $n=-0.226 $, $\xi_{o}=1$.

The per-capita ratios $a_{k}$ plus the exponent $n$ appearing in the 
scaling of Eq.(\ref{eq:mu}) are the minimum ingredients required
to derive a complete description of log-periodic corrections to
scaling and finite-time singularities within the framework of an
stochastics theory based on birth-death clustering processes.  The
positive state distributions $p_{j}$ are determined by $n$
and $a_{k}$ which also relate the exponent $\beta$, $C$ and $\psi$
as in Eq.(\ref{eq:gt5}) and $t_{o}$ as in Eq.(\ref{eq:to}).  This means 
that such state distributions of the system drive the log-periodic oscillations.  
We believe this feature of our stochastic model can help to elaborate a general
microscopic theory to understand the underlying mechanisms of
log-periodic patterns.  In the case of financial systems, such
microscopic theory should also explain the peculiar statistical
features in short time scales such as the highly correlated
variance or volatility of price fluctuations \cite{Bou98,Sor00},
by exploring the state distribution coefficient given by Eq.(\ref{eq:coeff}).

\newpage

\section*{Appendix: The Galerkin formulation}

In this appendix, the key features of the standard Galerkin 
finite elements method are stated concisely for completeness \cite{Fle84}.  
If a 2D problem in a domain $D(x,y)$ is governed by a 
linear differential equation $L(u)=0$, with boundary conditions $S(u)=0$ 
on $\delta D$, {\it i.e.}, the boundary of $D$.  Then, the Galerkin method
assumes that $u$ can be accurately represented by the approximate trial 
solution
\begin{equation}\label{eq:ape1}
u(x,y) = u_{o}(x,y) + \sum_{j=1}^{N} a_{j}(y)\phi_{j}(x) \;\;\;
\end{equation}
where the $\phi_{j}$'s are known, trial analytical functions, $u_{o}$ 
is chosen to satisfy the boundary conditions, and the $a_{j}$'s are test functions 
to be determined.

To obtain the unknown $a_{j}$'s, the inner product of the weighted residual $R$
is set equal to zero:
\begin{equation}
(R,\phi_{k}) \equiv \int\int_{D} R \phi_{k} \; dx dy =0 \;\;\; , \;\;\; k=1, \cdots, N \;\;\; ,
\end{equation}
where
\begin{equation}
R(a_{0}, a_{1} \cdots a_{N}, x, y) \equiv L(u) = L(u_{o}) + \sum_{j=1}^{N} 
        a_{j}(y)L(\phi_{j}) \;\;\; .
\end{equation}

Since this example is based on a linear $L(u)$, then the above can be rewritten as a matrix
equation for the $a_{j}$'s as
\begin{equation}
\sum_{j=1}^{N}a_{j}(t)L(\phi_{j},\phi_{k}) = - L(u_{o},\phi_{k}) \;\;\; .
\end{equation}

Substitution of the $a_{j}$'s resulting from this equation into Eq.(\ref{eq:ape1})
gives the required approximate solution $u(x,y)$.

\newpage
 
\section*{Figure caption}

\begin{itemize}
\item {\bf Fig.1}: Illustrative examples of log-periodic patterns.
Full line curves are the fit of our birth-death clustering theory
using Eq.(\ref{eq:gt5}).  These fits allow to estimate the total 
sum of outcomes $\lambda_{k}$.
\end{itemize}

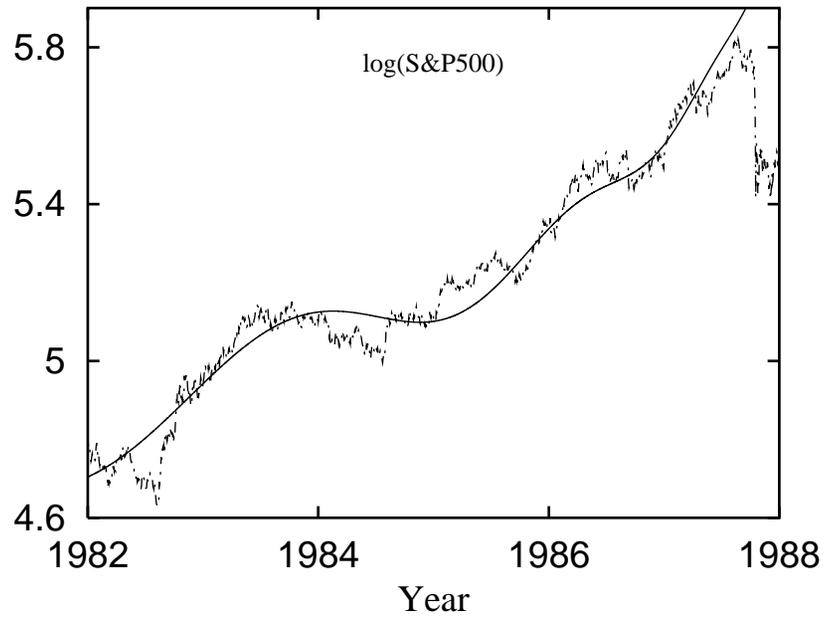
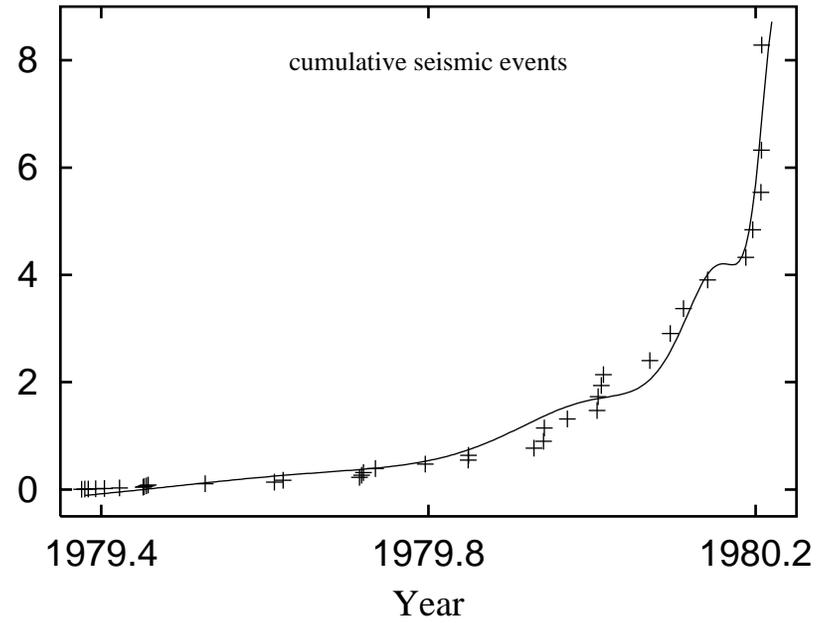
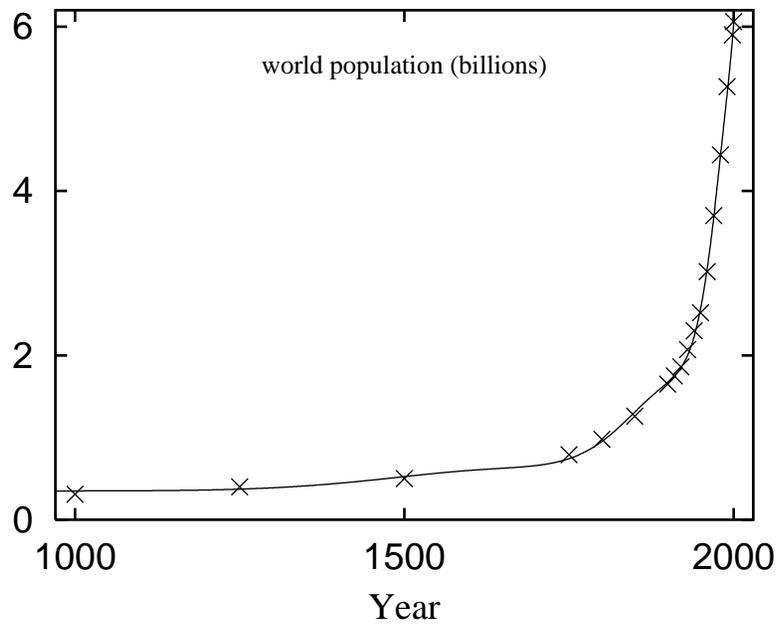
\end{document}